\documentclass[12pt,preprint]{aastex}

\newcommand{\nh}{N_{\rm H}}

\newcommand{\pica}{{\it Pictor A~}}
\newcommand{\Pic}{Pictor A~}

\newcommand{\xmm}{{\it XMM-Newton~}}
\newcommand{\chandra}{{\it Chandra~}}

\newcommand{\Bic}{{\it B$_{IC}$~}}
\newcommand{\Beq}{{\it B$_{eq}$~}}
\newcommand{\ke}{{\it k$_{e}$~}}
\newcommand{\ue}{{\it u$_{e+p}$~}}
\newcommand{\um}{{\it u$_{m}$~}}
\newcommand{\gmin}{$\gamma_{min}$}
\newcommand{\E}{\textbf{E~}}
\newcommand{\W}{\textbf{W~}}
\newcommand{\euno}{\textbf{e1~}}
\newcommand{\edue}{\textbf{e2~}}
\newcommand{\etre}{\textbf{e3~}}

\newcommand{\wuno}{\textbf{w1~}}
\newcommand{\wdue}{\textbf{w2~}}
\newcommand{\wtre}{\textbf{w3~}}
\def \nh {N${\rm _H}$~}
\shorttitle{Pictor A: X-ray mapping of the radio lobe}
\shortauthors{Migliori G. et al.}
\begin{document}
\title{Radio Lobes of Pictor A: an X-ray spatially resolved Study}
\author{Giulia Migliori}
\affil{Dipartimento di Astronomia, Universit\`a di Bologna,
via Ranzani 1, 40127 Bologna, Italy\\
Istituto di Astrofisica e Fisica Cosmica-Bologna, INAF,
 Via Gobetti 101, I-40129 Bologna,Italy\\
SISSA/ISAS, Via Beirut 2-4, I-34014, Trieste, Italy}
\email{migliori@iasfbo.inaf.it}
\author{Paola Grandi}
\affil{Istituto di Astrofisica e Fisica Cosmica-Bologna, INAF,
 Via Gobetti 101, I-40129 Bologna,Italy}
 \email{grandi@iasfbo.inaf.it}
\author{Giorgio G.C. Palumbo}
\affil{Dipartimento di Astronomia, Universit\`a di Bologna,
via Ranzani 1, 40127 Bologna, Italy}
\email{giorgio.palumbo@unibo.it}
\author{Gianfranco Brunetti}
\affil{Istituto di Radioastronomia-Bologna,INAF, Via Gobetti 101, 
I-40129 Bologna,Italy,}
\email{g.brunetti@ira.inaf.it}
\author{Carlo Stanghellini}
\affil{Istituto di Radioastronomia-Bologna,INAF, Via Gobetti 101, I-40129 Bologna,Italy,}
\email{c.stanghellini@ira.inaf.it}

\begin{abstract}
A new \xmm observation has made possible a detailed study of both lobes of 
the radio galaxy Pictor A. Their X-ray emission is of non thermal origin and 
due to Inverse Compton scattering of the microwave 
background photons by relativistic electrons in the lobes, as previously 
found. 
In both lobes, the equipartition magnetic field (\Beq) is bigger
than the Inverse Compton value (\Bic), calculated from the radio and X-ray 
flux ratio.
The \Beq/\Bic ratio never gets below 2, in spite of the large number of 
reasonable assumptions tested to calculate \Beq, suggesting 
a lobe energetic dominated by particles. 

The X-ray data quality is good enough to allow  a spatially resolved 
analysis. Our study shows that  \Bic varies through the lobes. 
It appears to increase  
behind the hot spots.
On the contrary, a rather uniform distribution of the particles is observed. 
As a consequence, the radio flux density variation along the lobes appears 
to be mainly driven by magnetic field changes.

\end{abstract}

\keywords{galaxies:individual (Pictor A) - radiation mechanisms: non-thermal -
  X-rays: galaxies}

\section{Introduction}
 
With the advent of the space observatories, \chandra and \xmm, fundamental progress in the study of 
X-ray emission from active galactic nuclei (AGNs) has become possible. X-ray spatially resolved studies 
of radio loud AGNs have been performed and those radiative 
processes, which are responsible for X-ray 
emission far from the nucleus, in the jets, hot spots and lobes, have been investigated.\\ 
In the last years many studies have been produced on each of these
structures ( Harris \& Krawczynski 2000, Hardcastle et al. 2002, Croston et
al. 2005, Kataoka \ Stawarz 2005).  
The aim is to reach a clearer knowledge of the
physics powering each component and of their relative  interplay.
Among the extended structures which emit X-ray photons, the lobes are the faintest and  most difficult 
regions to detect.  On the other hand, X-ray observations provide important physical information. 
In fact, it is thought that X-ray emission in the lobes is due 
to Inverse Compton (IC) by relativistic electrons.
IC scattering in radio lobes can occur either 
between $\gamma \approx 10^3$ electrons and CMB photons
(Harris \& Grindlay 1979, Miley 1980) or 
 between $\gamma \approx 100-300$ electrons and 
nuclear photons (Brunetti et al. 1997, 2000). The last case is more relevant in the context of powerful and compact ($<
100$ kpc) radio galaxies.
A comparison  of the radio and X-ray fluxes provides a direct estimate of 
the average magnetic field (B$_{IC}$) along the line of
sight and, independently the number densities of the 
 IC emitting particles in the lobes. 

Two are the main concerns in this kind of study: good statistics (radio lobes
are weak X-ray sources) and spatial resolution to distinguish 
non thermal from thermal emission.
In fact it is possible that external gas, compressed by the expansion of the radio lobes, 
emits in the X-ray band by thermal bremsstrahlung (Kraft et al. 2000, Reynolds et al. 2001)
Obviously the task is even more difficult if the radio galaxy is
situated in a cluster.    
In spite of these problems, X-ray detections of extended synchrotron
radio regions were already been achieved focusing on low redshift radio
galaxies (Feigelson et al. 1995, Kaneda et al. 1995, Brunetti et al.
1999, Tashiro et al. 2001).  
These early results from X-ray studies on lobes 
have opened a new field of research and in particular have allowed the first
checks on the assumption of minimum energy condition in the lobe (equipartition). 
Several detections of Radio Lobes have been successively performed with Chandra
and XMM-newton (Isobe et al. 2002, Grandi et al. 2003, Comastri et al. 2003, Hardcastle et
al. 2005, Croston et al. 2005, Kakatoa $\&$ Stawarz 2005) and the
equipartition condition has been further investigated. There is not yet a general
agreement: in some cases a substantial concordance between
equipartition and Inverse Compton magnetic field has been claimed 
(Brunetti et al. 2001, Hardcastle et al. 2002, Belsole et al. 2004, 
Croston et al. 2005, 
Overzier et al. 2005), in other cases a moderate
violation of equipartition (\Beq $>$
\Bic) has been found (Brunetti et
al. 1999, Isobe et al. 2002, Comastri et al. 2003, Grandi
et al. 2003, Kataoka et al 2003,Bondi et al. 2004).
It is clear that slightly different assumptions drive to the two
opposite results. The most reasonable conclusion is that the unbalance 
between the particle and magnetic field
energy density is not dramatic and the equipartition argument 
is a viable zero-order reference guide.

\noindent
The spatial distribution of both magnetic field and particles within the
lobes themselves is a second connected branch of
investigation. Specifically studying the variations of magnetic fields and
particle densities with position turns out to be a useful procedure to get
information on the dynamical plasma evolution.
Observationally attempts in addressing these problems have been performed for
only a few Radio Galaxies. The results indicate an 
enhancement of the magnetic energy density towards the edge of the lobes
(Tashiro et al. 01, Isobe 2002, 2005 )
   

\Pic, a famous nearby (z=0.035) radio bright ($P_{408MHz}=6.7\times10^{26}$ W
Hz$^{-1}$) Fanaroff Riley II (FRII) radio galaxy, is an excellent candidate for studying  
X-ray IC emission from the lobes.  
Actually, \xmm discovered X-ray radiation associated to the extended radio
lobes (Grandi et al. 2003). 
In spite of the relatively  short exposure time ($\sim 15$ ks) and the very
high background, a reasonably good spectrum was obtained from a circular
region centered on the east lobe. 
However, the data interpretation was not univocal, as the X-ray spectrum could
be described 
by both a power law ($\alpha_{X}=0.6\pm 0.2$) and a thermal model ($kT\sim 5$ keV). 
The presence  of an extended emission, most likely to be ascribed to non-thermal processes, has been recently confirmed (Hardcastle \& Croston 2005). 

In order to definitively solve the ambiguity about the nature of the radio-lobe X-ray emission, 
\xmm re-pointed \Pic on January 14, 2005.  The observation was successful and, 
beyond our expectation, the MOS1 signal-to-noise ratio was good enough to allow a spatially resolved 
spectral analysis of both the west and east radio lobes.

\section{\bf Data Reduction}
 
The radio lobes were the target of the new \xmm observation. For this reason, the EPIC/p-n and EPIC/MOS1 cameras 
operated in full-frame mode, with thin optical filter. The EPIC/MOS2 camera operated in small-window 
mode in order to assure the study of the nucleus without any pile up problem.
P-n data were excluded  because of a substantial problem with nuclear PSF, which covers a great fraction of 
the X-ray emission from the lobes. Therefore our study is only based on the MOS1 instrument data.

Data reduction was performed using the software package SAS, Version 6.0. A
calibration index file appropriate 
for the date of the observation and data analysis (July 2005) was produced. We identified two high background periods (background flares) and  we excluded all periods with a count rate higher than 0.35 cts s$^{-1}$.
After the cleaning procedure, the net total exposure time was $\sim 50$ ks.
We selected only those events with {\bf PATTERN $\leq$ 12} (single, double, triple and quadruple events) and imposed 
the  filter {\bf \#XMMEA\_EM} in order to exclude artifact events.

\subsection{\bf Selection of the studied regions}
\begin{figure}[h]
 \centering
\includegraphics[width=8.5cm, angle=-90]{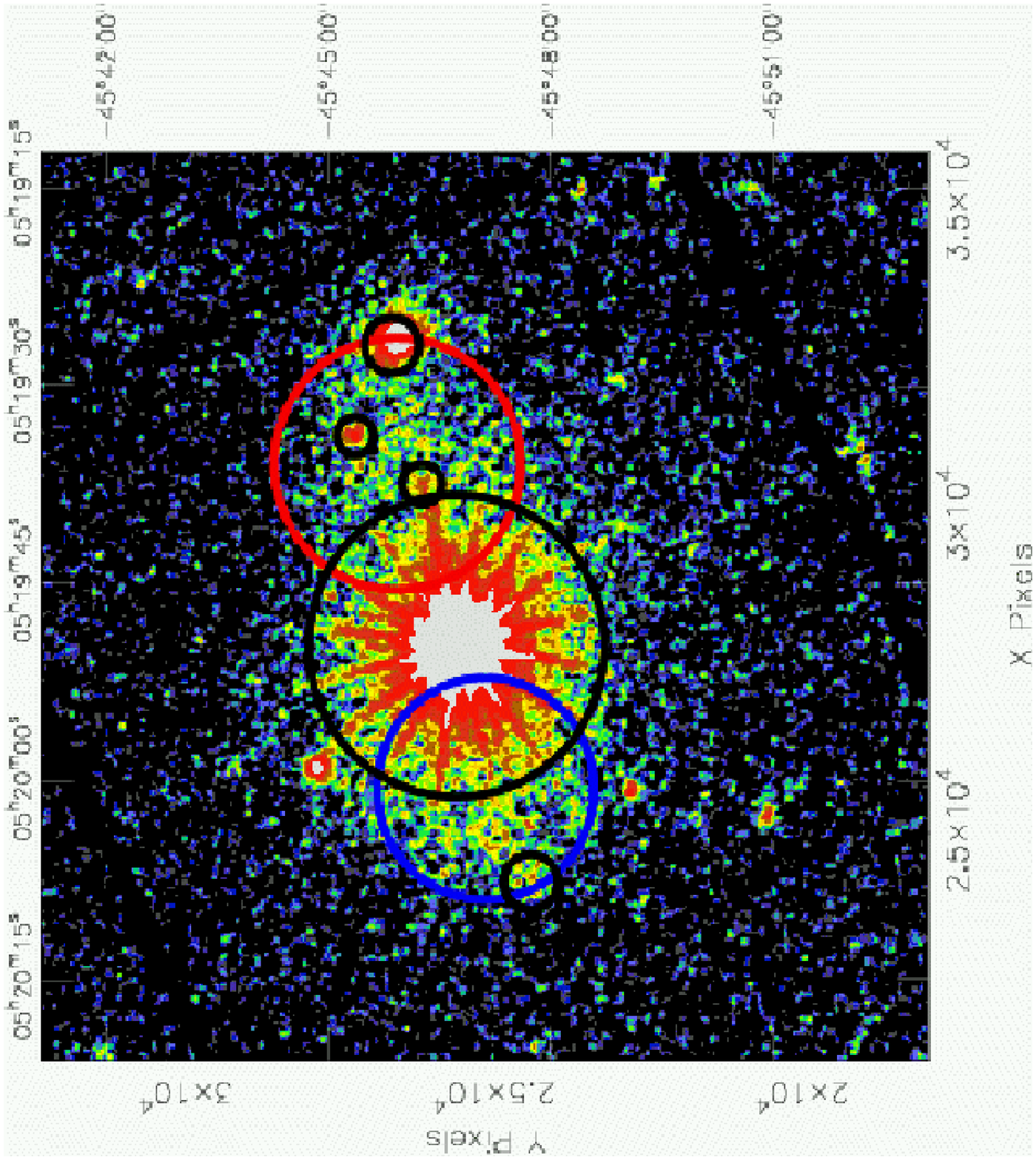}
\includegraphics[width=8.5cm, angle=0]{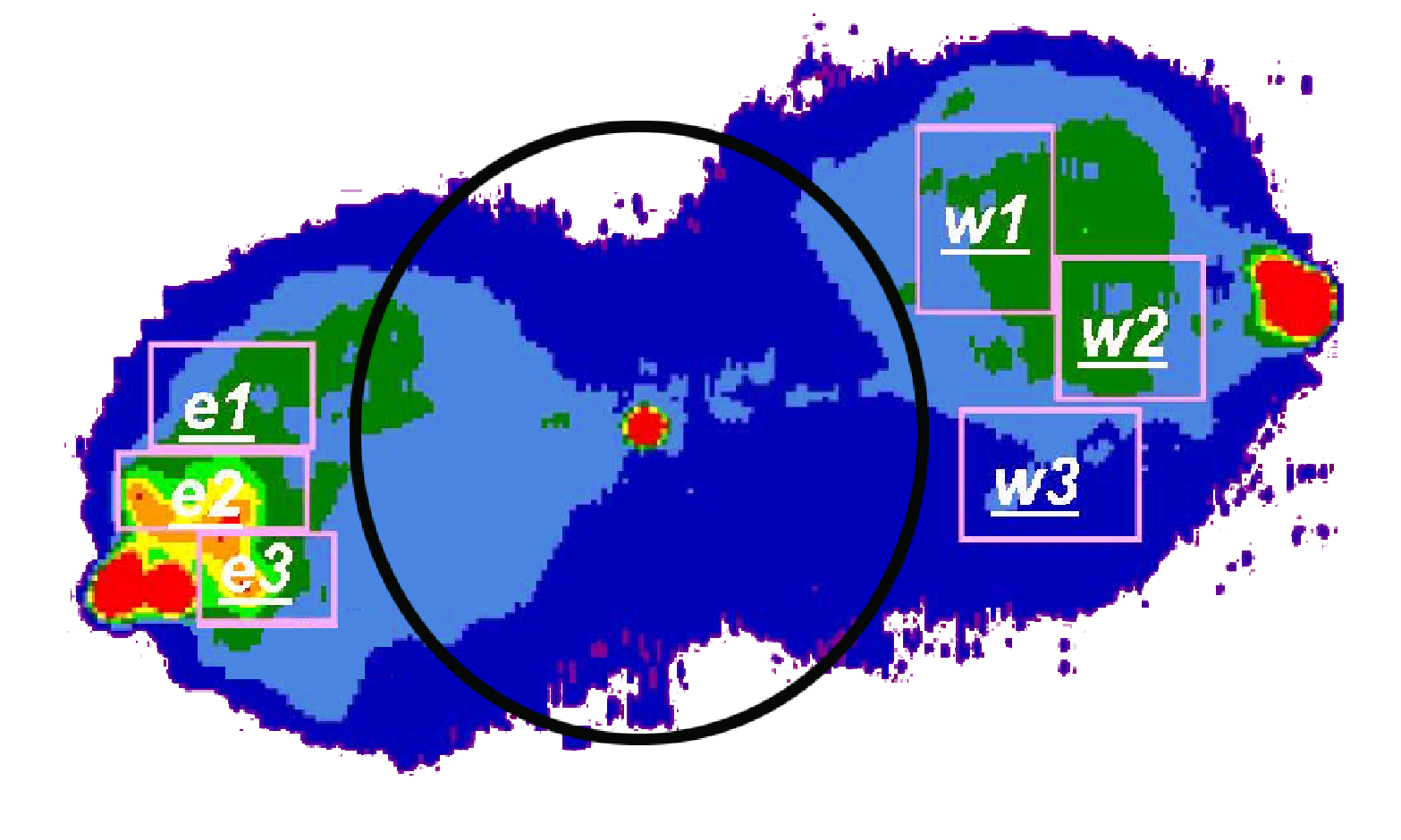}
\caption{{\it (Upper Panel)} -- \xmm/MOS1 image (0.2-10 keV) of Pictor A observed
on 14 January 2005. Several components are visible: the bright nucleus, the west
hot spot, the jet and the two lobes. 
The blue and  red circles represent the east (\E) and west (\W) extraction  regions of the lobes, respectively. 
Regions (black circles) corresponding to the two hot spots, jet contribution and a point source, are excluded. 
{\it (Lower Panel)} --  VLA map (Perley et al. 1997) at 20 cm. Pink boxes represent the sub-regions 
used for the spatially resolved analysis. Labels described in the text. Black circle delimitates the excluded nuclear region.}
\label{PicA_4}
\end{figure}

VLA observations (Perley et al. 1997) of \Pic show two nearly
circular radio lobes with hot spots and a faint radio jet connecting the
nucleus to the west hot spot. The same features can be observed in Figure \ref{PicA_4} ({\it Upper Panel}), where 
the MOS1 image (0.2-10 keV) is shown. 
If the 20 cm radio contours are superposed on the \xmm image, the spatial
coincidence between radio and X-ray emission is unambiguous. In particular, 
the X-ray counterparts of the radio lobes are clearly visible as elongated emission around the bright nucleus.
In comparison with the previous \xmm observation, performed in 2001 (Grandi et al. 2003), the good quality of this new 
data has allowed two different kinds of analysis. In fact, we  studied   each
lobe as a whole, but also performed a {\it spatially resolved individual
  study} on both lobes. In the following, we give a description of the regions 
chosen for the two studies.\\
\noindent 
{\bf The extended regions: E and W} -- The selection of the studied regions was based on both X-ray and radio images. 
We selected two circular regions, one for each lobe, with different radius (r)
(r=88'' for the east lobe  and r=100'' for the west lobe). Afterwards we will
refer to  these two regions as  \E  and \W  for east and west lobes, respectively.
In order to avoid any nuclear contamination we excluded the  emission coming  from a circular region of r=120'' centered on the nucleus. This choice is also supported by considerations based on the calibration report by Ghizzardi (XMM-SOC-CAL-TN-0022). She clearly shows that 80\% of the nuclear source emission is contained within about 30 arcsec. Moreover, following Ghizzardi's prescriptions we have evaluated encircled energy fraction for the nucleus in the case of an annulus comprising our selected lobe regions (R$_{min}$=120'', R$_{max}$=222''. The result is that any possible nuclear contamination should be of the order of the measured errors for the fluxes.\\
In the case of the west lobe, it was also necessary to exclude the jet
emission (a circle with r=22'') and  a point-like source (r=15''). The
excluded regions are represented in Figure \ref{PicA_4} ({\it upper panel}).

\noindent
{\bf The subregions} -- In order to perform a spatially resolved study of the
lobes, we selected rectangular 
subregions in each lobe, as shown in Figure \ref{PicA_4} ({\it lower pane}). In what follows, we indicate
these regions as  \euno, \edue, \etre for the east lobe, and \wuno,
\wdue, \wtre for the west lobe, counting from top to bottom.

\subsection{\bf X-ray Spectral Analysis}

We accumulated spectra of the two regions \E and \W,  and of the subregions
previously described, i.e.  \euno, \edue, \etre and \wuno, \wdue, \wtre in the
energy range 0.2-10 keV. In each case, the background was extracted from a
region of the same extension taken on the same CCD. \\ 
Spectral analysis was conducted on two different levels. 
First, we maximized the statistics and considered
the \E and \W spectra in order 
to define which radiative process is causing the X-ray emission. 
Afterwards, we  analyzed  the sub-region spectra. Spectral analysis has been
done with XSPEC, 
Version 11.2.0. Every spectrum was grouped in order to ensure a minimum of 25 net counts in each bin.
Throughout the paper, all errors are quoted at 90\% confidence for two interesting parameters ($\Delta\chi^2=4.61$).

\subsection{\bf Radio Fluxes}

The radio fluxes of all the analyzed regions were derived at 20 cm
from the VLA
image produced by  Perley et al. (1997).
Uncertainties in measurements of flux density of extended structures are 
difficult to
determine for interferometric data. Extended emission may be lost 
because the
corresponding spacial frequencies are not sampled by the array of antennas.
Furthermore this radio source is at low declination and the VLA observed it
at a very low elevation, with not negligible atmospheric opacity at this 
frequency.
These effects have been taken in consideration by Perley et al.
who find the VLA total flux density, after the standard calibration,
of about 6$\%$ lower respect to the total flux
density derived by single dish measurements. They ascribe this difference
to the atmospheric opacity only, and correct the image accordingly.
Given the considerations above the uncertainty in the flux density
measurements is dominated by residual calibration errors which can be 
estimate to be about 3$\%$. 

\section{\bf E and W regions}
\subsection {\it Thermal versus non-thermal emission} 
The first step of the study was aimed at solving  the 
ambiguity (thermal vs  non-thermal) about the  origin of the lobes X-ray
emission.  Consequently, we tested two different models: a thermal emission
from hot diffuse gas ({\it mekal} model in XSPEC, Mewe et al. 1985) and  a
non-thermal radiation  modeled with a pure power law.
In both cases, the acceptable range of {\nh} is consistent with the Galactic
line-of-sight value {\nh}$^{Gal}=4.18\times10^{20}$ cm$^{-2}$. Thus we fixed the
column density to the Galactic value in order to reduce the parameters uncertainties. 
Results in Table 1 strongly suggest a non-thermal origin of the radiation. In each lobe
a power law is preferred to a thermal model and the X-ray photon
indeces are in very good agreement with the average radio slopes 
($\alpha_{R} \sim0.8$) measured by Perley et al. (1997).
\begin{table*} 
\begin{flushleft}
\begin{center}
\begin{scriptsize}

\caption[] {XMM-Newton fit to the Western \W and Eastern \E Lobe of \pica in the
  0.5-10 keV band. \nh is fixed to the galactic value \nh$^{Gal}$= 4.18$\times10^{20}$
  cm$^{-2}$}
\begin{tabular}{lcccccc}
\noalign {\hrule}


&&&&&&\\
&\multicolumn{3}{c}{Power Law} & \multicolumn{3}{c}{Thermal Emission}\\
&&&&&&\\

              &$\Gamma$    &$\chi^2$(dof)  &Flux$^a$     &kT$^b$    &$\chi^2$(dof)  &Flux\\
              &            &               &(0.5-2  keV)  &(keV) &               &(0.5-2  keV)\\
&&&&&&\\
\hline\\ 
\W             &1.7$^{+0.2}_{-0.2}$ &29(38)  &$12\pm1\times10^{-14}$  &$6^{+3}_{-1}$  &39(38)   &$12\pm1\times10^{-14}$\\
&&&&&&\\
\E             &1.8$^{+0.2}_{-0.2}$ &33(31)  &$9\pm1\times10^{-14}$ &$5^{+3}_{-2}$ &45(31)  &$8\pm1\times10^{-14}$\\
&&&&&&\\
\hline
&&&&&&\\
\multicolumn{7}{l}{$^a$ - X-ray Flux corrected for Galactic Absorption}\\
\multicolumn{7}{l}{$^b$ - Metal abundance fixed at 0.5}\\
\end{tabular}
\end{scriptsize}
\end{center}
\end{flushleft}
\end{table*}
\noindent
The non-thermal X-ray flux of the east lobe is consistent with the previously
reported one by Grandi et al. (2003), once it is rescaled to their (smaller) extraction area.
However, to completely explore all the possibilities, a mixed of thermal ({\it
mekal}) and non-thermal (power law) radiation was also tested to
verify the presence of gas within the lobes. 
The dual component model does not significantly improve the fit. 
Even if present, the thermal component is negligible and
its  contribution to the the total $0.5-2$ keV  flux is not larger than 15$\%$.

\subsection{\it Lobes Energetic} 

Having  established the non-thermal nature of the X-ray emission, 
we could directly determine the magnetic field (\Bic) in both lobes
by assuming that the emission is due to IC scattering of
CMB photons.
If, the synchrotron  radio ($L_{syn}$) and Compton scattered X-ray ($L_{IC}$)
luminosities are known (e.g., Blumenthal \& Gould 1970)
\begin{center}
\begin{equation}
L_{syn}=C_{syn} k_e V B^{\alpha+1} \nu^{-\alpha} 
\end{equation}
\end{center}
and 
\begin{equation}
L_{IC}=C_{IC} k_e V \nu^{-\alpha}
\end{equation}
\noindent
\Bic can be immediately estimated 
\begin{equation} 
B_{IC} = [\frac{F_{1.4~
    GHz}}{F_{1~keV}}\frac{C_{IC}(\alpha) (1+z)^{\alpha+3}}{C_{sin}(\alpha)}]^{\frac{1}{\alpha+1}}[\frac{\nu_{syn}}{\nu_{IC}}]^{\frac{\alpha}{\alpha+1}}.
\end{equation}
\noindent
$C_{IC}$ and $C_{syn}$ are quantities depending only from
$\alpha$ and  $V=A\times s$  is the volume, A being the area (arcsec$^2$) 
projected on the sky plane  and s the thickness of the analyzed region.
$k_e$ is the electron density having assumed a power law distribution for the 
electrons $N(\gamma) =k_e \gamma^{-(2 \alpha+1)}$ and
$\alpha=\Gamma-1=\alpha_x=\alpha_R$. 

\ke can be derived from the synchrotron luminosity (1), 
and thus the lobe energetics, i.e. magnetic field and particles energy  
densities (respectively \um and \ue) can be estimated
\begin{equation} 
u_{m} =\frac{B_{IC}^2}{8\pi}
\end{equation}
\begin{equation} 
 u_{e+p}= \frac{(1+k)mc^2}{2\alpha-1}k_e (\gamma_{min})^{1-2\alpha}.
\end{equation}
\noindent
The ratio between electron and proton densities $k$ was set equal to 1 and the low energy cut-off of the electron spectrum \gmin $=50$. We also assumed
$\alpha=0.8$, as indicated by both radio and X-ray observations.





The results (Table 2) show very similar
physical conditions in the two lobes (\E and \W regions). 
The strength of the magnetic field  ($B_{IC} \sim 3\mu G$) is the same in 
both lobes and no significant difference is found 
for the particle density \ke.
Most relevant is the dominance of the particle energy density on 
the magnetic field energy density. Energetic of both lobes appears 
to be dominated by particles 
(\ue/\um=$(56 \pm 10) (\gamma_{min}/50)^{-0.6}$ 
for \E and \ue/\um=$(50 \pm 10) (\gamma_{min}/50)^{-0.6}$
for \W), confirming  a possible departure from  the equipartition
principle.\\ 
We are aware that this is a very delicate point, since a large number of
assumptions are necessary to estimate the lobe energetics.
In Appendix we discuss the dependence on parameters, $k$ \gmin and $s$, 
and show that the ratio between the equipartition and IC magnetic
fields, \Beq/\Bic, when we consider plausible variations of these parameters, never goes below a factor of 2.
Even considering a more complex particle distribution (a broken power law) the \Bic and \Beq
discrepancy can not be reduced to a value less than $\sim1.6$ (Hardcastle \&
Croston 2005).

\section{\bf Spatially resolved Analysis}

The analysis performed on the whole lobes was successively extended to 
the sub-regions data set. Data of each subregion were fitted with a power 
law  absorbed by Galactic \nh. 
Since the low number of counts (ranging from a minimum of 125 to a maximum of
216), the X-ray spectral slopes were poorly constrained and therefore 
we decided to fix $\Gamma=1.8$ in each sub-region.
This choice allowed to better constrain the X-ray fluxes and to map 
physical quantities (namely magnetic field and particles) through the lobes, 
following the same procedure adopted in the previous section. 
The results are in Table 2.
\begin{table*}[h]
\begin{center} 
\begin{flushleft}
\begin{scriptsize}
\caption[]{Magnetic field and particles energy  densities of the  Western \W and Eastern \E Lobes and Spatially resolved Analysis Results:}
\begin{tabular}{lccccccc}
&&&&&&&\\
\hline
&&&&&&&\\
      &Flux$_{1.4GHz}$        &Flux$_{\rm 0.5-2 keV}$  & Area          &\Bic      &\ke      &\ue/\um    &\ue\\
            & (Jy)           &(erg cm$^{-2}$ s$^{-1}$) & (arcsec$^{2}$)              &($\mu$G)    &($\times10^{-5}$cm$^{-3}$ ) 
&    & ($10^{-11}$ erg cm$^{-3}$)\\
&&&&&\\
\hline\\
&&&&&&&\\
\multicolumn{8}{c}{\bf East lobe}\\
&&&&&&&\\
\E       &$11.4\pm0.3$       &$9.0\pm1.0\times10^{-14}$       &$12776\pm383$     &$3.1\pm0.2$         &$8.2\pm1.0$     &$56\pm10$                         &$2.1\pm0.3$\\
&&&&&&&\\
\euno      &$2.5\pm0.1$  &$1.9\pm0.3\times10^{-14}$ &$2992\pm90$    &$3.3\pm0.3$         &$7.3\pm1.0$    &$44\pm11$                &$1.9\pm0.3$\\

&&&&&&&\\
\edue      &$2.8\pm0.1$             &$1.2\pm0.2\times10^{-14}$                   &$1824\pm55$     &$4.4\pm0.5$         &$7.5\pm1.6$    &$25\pm8$                 &$2.0\pm0.4$\\
&&&&&&&\\
\etre      &$2.5\pm0.1$             &$1.5\pm0.2\times10^{-14}$                   &$1872\pm56$      &$3.5\pm0.3$         &$10.2\pm1.6$   &$55\pm12$                &$2.7\pm0.4$\\
&&&&&&&\\
\hline\\
&&&&&&&\\
\multicolumn{8}{c}{\bf West lobe}\\
&&&&&&&\\
\W         &$13.7\pm0.4$           &$12.0\pm1.0\times10^{-14}$   &$21408\pm642$     &$2.9\pm0.2$         &$6.5\pm0.8$     &$50\pm10$                         &$1.7\pm0.2$\\
&&&&&&&\\
\wuno      &$3.3\pm0.1$             &$2.4\pm0.3\times10^{-14}$               &$3640\pm109$&$3.2\pm0.3$         &$8.0\pm1.4$    &$51\pm13$                &$2.1\pm0.4$\\      
&&&&&&\\
\wdue      &$2.8\pm0.1$             &$2.4\pm0.3\times10^{-14}$               &$3024\pm91$ &$3.2\pm0.3$         &$8.1\pm1.4$    &$52\pm13$                &$2.1\pm0.4$\\
      
&&&&&&&\\
\wtre      &$1.5\pm0.1$             &$2.9\pm0.5\times10^{-14}$               &$3400\pm102$  &$2.5\pm0.2$         &$7.7\pm1.1$    &$81\pm17$                &$2.0\pm0.3$\\    
&&&&&&\\
\hline
\multicolumn{8}{l}{}\\
\multicolumn{8}{l}{$^a$ -- Radio flux uncertainties of $3\%$ reflect
  calibration uncertainties of the antenna}\\
\multicolumn{8}{l}{$^b$ --Areas of each sub-region are measured using the 20 cm
  radio map. See also Fig. 1 ({\it Lower Panel})}\\
\end{tabular}
\label{fluxes}
\end{scriptsize}
\end{flushleft}
\end{center}
\end{table*}
\noindent
In spite of the quite large uncertainties, Table 2 suggests
that  
\Bic increases behind the east hot spot, where the radio flux is higher
(see {\it Lower Panel} of Fig. 2).\\
On the contrary, no change of the electron density appears evident.

\begin{figure}
 \centering
\includegraphics[width=10cm]{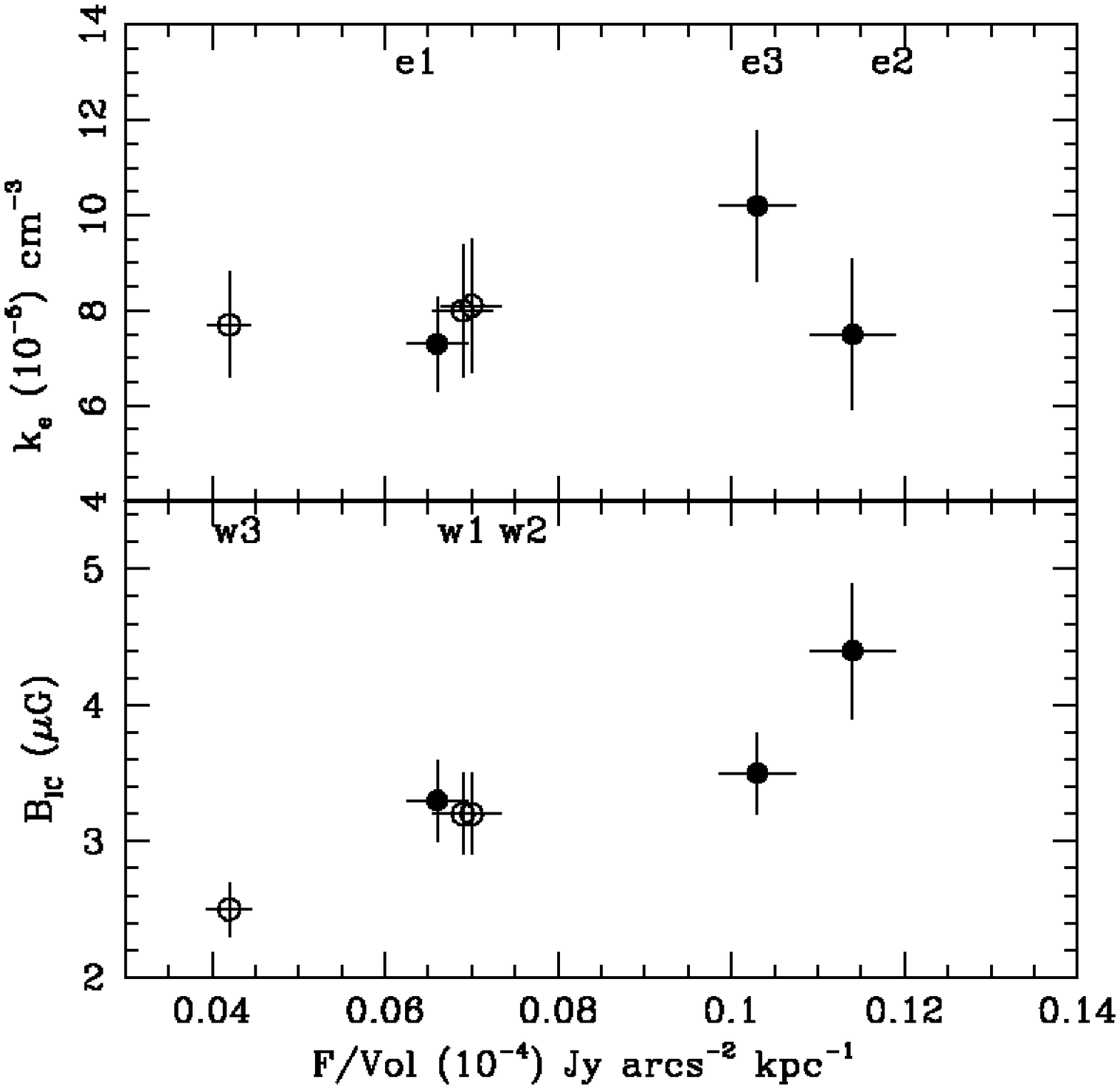}
\caption{ \Bic ({\it Lower Panel}) and \ke ({\it Upper Panel}) values of east (solid circles) and west (empty circles) sub-regions are plotted as a
 function of the radio flux ($F/V$) at 1.4 GHz, normalized by the relative volume.}
\label{flubic}
\end{figure}
\noindent
Combining data from both lobes, the previous results are strengthen.
In Figure \ref{flubic}, \Bic and \ke are plotted as a function of radio
flux (divided for the corresponding sub-region volume).  \Bic variation is statistically significant 
($\chi^{2}=4.7$ with a probability $p=3.0 \times 10^{-4}$) and traces the
variation of the radio flux density. 
A {\it Spearman test} gives a correlation coefficient $r=0.8$ with a $s=0.0499$
significance.\\
Again no trend is found between $k_e$ and the radio flux in Figure \ref{flubic}
(upper panel). The correlation test gives $r=0.31$ with $s=0.54$.\\
We note that these conclusions are based on the hyphotesis that the spectral
energy distribution of the particles does not vary through the lobe, i.e. that
our assumption of $\Gamma - 1= \alpha$ is valid for each sub-region analyzed.
This is a very crucial point, as in principle 
spectral index variations among the 
analyzed sub-regions can affect the \Bic -- radio flux trend
(observed in Figure 2).
In order to exclude this possibility, we analyzed the spectral index radio
maps between 20 and 6 cm and between 6 and 2 cm (kindly provided by
R. Perley and shown in Fig. 5 of Perley et al. 1997).
The maps shows negligible spectral index variations ($\Delta \alpha \leq
0.05$) from a sub-region to another one, confirming the robustness of our
result.

The different behaviour of \Bic and $k_e$  might indicate a
dynamical  decoupling between the magnetic field and the electron energy
densities (i.e changes of the \ue$/$\um ratio along the lobes) as found 
in 3C452 by  Isobe et al. (2002, 2005).
It is  possible that the plasma expands along the regions immediately 
beyond the hot spots, while  (high energy) electrons diffuse away on larger scales providing a 
fairly constant density distribution.
An efficient spatial diffusion of the high energy electrons
in the lobes and  a decreasing magnetic field with distance from hot
spots (yielding particles decoupling)
was suggested by Blundell \& Rawlings (2000) to explain the discrepancy 
between the radiative lifetimes of electrons and dynamical ages of double
radio galaxies. In this case 
if relativistic energetic particles stream through 
the plasma replenishing the lobes, the synchrotron
spectral steepening are driven by the magnetic field decay 
during the plasma expansion rather than the emitting
particles aging. 
Our results in Fig.2 are in agreement with this picture, which predicts an increasing $u_{e+p}/u_B$ ratio with the hot spot distance.

\section{Conclusions}

A new XMM-Newton observation has solved the problem about the origin of the
X-ray emission from the lobes of Pictor A. 
X-ray emission is due to IC scattering of CMB photons, confirming the study
based on previous, suggestive but still uncertain, observations (Grandi et
al. 2003, Hardcastle \& Croston 2005, Kataoka \& Stawarz 2005).\\ 
The IC magnetic field, estimated for each lobe, is a factor of 2.7 
below equipartition value. 
A significant variation of key parameters, in the equipartition
formula, can not account for this discrepancy: energetics of the lobes appear to
be dominated by particles.\\ 
The good quality of the new observation has permitted a spatially resolved
study of the lobes. Interesting results emerge from the study of the east
lobe. In this case, an indication of an increase of magnetic field is registered in the region
behind the east hotspot. A similar behaviour of the magnetic field is also considered in the model proposed by Hardcastle et al. (2005).\\
 Moreover, when we combine  the results 
of the two lobes,
there is evidence for a correspondence between magnetic field and radio flux
density variations. On the other hand, we do not observe a similar  trend in
electron density. Our results suggest a quite uniform distribution of 
particles in the lobes against a  spatial variation of the magnetic field.
This is in line with the figure of a magnetic field 
which decreases with the increasing distance from the hot spot proposed 
by Blundell \& Rawlings (2000).



\begin{acknowledgements}
We are very grateful to R. Perley for kindly providing quantitatively analyzable
radio images of Pictor A. 
\end{acknowledgements}

\begin{appendix}
\section{Checking the Equipartition Condition in the lobes}

\noindent
In this Appendix we use 
the  equipartition formula revised by Brunetti et al. (1997) which accounts for a minimum energy of the emitting electrons, 
\begin{equation} 
B_{eq}=[C(\alpha)(1+k)\frac{L_{syn}(\nu)\nu^{\alpha}}{V}]^{\frac{1}{\alpha+3}}\gamma_{min}^{\frac{1-2\alpha}{\alpha+3}},
\end{equation}
where C($\alpha$) is a quantity depending only from $\alpha$, we calculate equipartition magnetic fields, \Beq, for all the \W and \E
lobes and compare them to the \Bic values estimated exploiting
the IC scattering of the microwave background radiation by relativistic 
electrons.
 It is also necessary to stress that the adopted formula differentiates from the traditional one (Pacholczyk, 1970) where magnetic field is calculated in the frequency band between 10 amd 100 MHz, corresponding to the frequency range observable with the radio telescopes.\\
At first,  we assume $\alpha=0.8$, $k=1$ and \gmin$=50$ and a  volume for
each lobe region V=Area (arcsec$^2$) $\times$ s where s is the path length, i.e
the thickness of the analyzed region (see Table \ref{fluxes}).
The path length $s$ is set equal to the lobe diameter ($s=95/h$ kpc) given by
Perley et al. (1997), implicity assuming a spherical geometry.
Errors on \Beq are deduced by  the propagation errors as done for \Bic, 
considering in this case the radio flux and Area uncertainties. 
Table \ref{BeqBic} clearly shows that, even if the \Beq and \Bic trend are very similar, 
their values are not consistent. The magnetic fields based on equipartition
condition overestimate \Bic by a factor $\sim 3$.\\
As it is well known, the influence of the unknown parameters can be 
important
on the final result of \Beq, however our tests show that reasonable 
variations of a 
single parameter does not 
cause important changes in the \Beq/\Bic ratio.
Indeed, the major uncertainty in the equipartition formula comes from our
ignorance of $\gamma_{min}$. Clearly in the case of Pictor A
$\gamma_{min}$ cannot be much greater than several $10^2$ as the synchrotron
emission detected from the west radio hot spot at 74 MHz (Perley et al. 1997), and 330 MHz comes from
$\gamma \approx 800-10^3$ electrons (Meisenheimer et al.~1997)
and adiabatic expansion from hot spots
to the lobes should reduce the energy of these particles. According to Blundell's calculations (Blundell et al. 2006), a reduction of a factor of 10 is expected for the Lorentz factors of all particles due to adiabatic expansion losses.
Anyhow in the case of Pictor A it is 
\Beq/\Bic$\propto \gamma_{min}^{-0.16}$ and even by adopting  very
conservative values $\gamma_{min} \approx 800- 10^3$ one would still
find \Beq/\Bic$ \approx 1.9- 1.7$.
Operating in the direction of a \Beq decreasing, we note that
in the case of the east lobe $k$=0 
(assuming the contribution of heavy particles to be null)
 gives \Beq/\Bic$ \simeq 2.2$, 
%
and 
$s=134$ kpc (accounting for projection effects for an inclination
of the radio axis of $\sim 45^o$)
gives \Beq/\Bic$ \simeq 2.6$.
Analogous results are found for the west lobe. 
Even if all three parameters are simultaneously forced to the minimum
value, we can not reach equipartition conditions.

\begin{table}[h]
\begin{center}
\scriptsize
\caption[]{Equipartition magnetic fields (\Beq)
, calculated with the revisein determinind formula by Brunetti et al.(1997), and their ratios with the relative \Bic. }
\begin{tabular}{ccc}
\hline
&&\\
        &\E     &\W   \\
&&\\
\Beq($\mu$G)    &$8.7\pm0.1$  &$8.8\pm0.1$  \\
&&\\
\Beq/\Bic       &$2.8\pm0.2$  &$2.7\pm0.2$\\
&&\\
\hline
\end {tabular}
\label{BeqBic}
\end{center}
\end{table}
\end{appendix}

\end{document}